\documentclass[twocolumn,aps,prc,preprintnumbers,showpacs,showkeys,nofootinbib,superscriptaddress,fleqn,floatfix,tightenlines, 10pt]{revtex4-1}
\usepackage{amsmath,amsfonts,amssymb,amscd,amsxtra,amsthm}
\usepackage{natbib}
\usepackage{graphicx,subfigure}  
\usepackage{epstopdf}
\usepackage{dcolumn}  
\usepackage{bm}          
\usepackage{slashed}
\usepackage[utf8]{inputenc}
\usepackage{CJK}
\usepackage{mathptmx}
\usepackage{mathrsfs}
\usepackage[normalem]{ulem} 
\usepackage[dvipsnames]{xcolor} 
\usepackage{array}
\usepackage{multirow}

\def\vc#1{\mbox{\boldmath $#1$}}

\begin{document}
\title{Possibility of generating the $^3_{\Lambda_c}$H in the quark-delocalization color-screening model}

\author{Siyu Wu}
\affiliation{Key Laboratory of Atomic and Subatomic Structure and Quantum Control (MOE), Institute of Quantum Matter, South China Normal University, Guangzhou}
\affiliation{Institute of Modern Physics, Chinese Academy of Sciences, Lanzhou 730000, China}
\affiliation{Guangdong Provincial Key Laboratory of Nuclear Science, Institute of Quantum Matter, South China Normal University, Guangzhou 510006, China}
\affiliation{Guangdong-Hong Kong Joint Laboratory of Quantum Matter, Southern Nuclear Science Computing Center, South China Normal University, Guangzhou 510006, China}

\author{Qian Wu}~\email{wuqian@smail.nju.edu.cn (co-first author)}
\affiliation{Institute of Modern Physics, Chinese Academy of Sciences, Lanzhou 730000, China}
\affiliation{Department of Physics, Nanjing University, Nanjing 210097, China}

\author{Hongxia Huang}~\email{hxhuang@njnu.edu.cn}
\affiliation{Department of Physics, Nanjing Normal University, Nanjing 210097, China}

\author{Xurong Chen}~\email{xchen@impcas.ac.cn}
\affiliation{Institute of Modern Physics, Chinese Academy of Sciences, Lanzhou 730000, China}
\affiliation{School of Nuclear Science and Technology, University of Chinese Academy of Sciences, Beijing 100049, China}

\author{Jialun Ping}~\email{jlping@njnu.edu.cn}
\affiliation{Department of Physics, Nanjing Normal University, Nanjing 210097, China}

\author{Qian Wang}~\email{qianwang@m.scnu.edu.cn}
\affiliation{Key Laboratory of Atomic and Subatomic Structure and Quantum Control (MOE), Institute of Quantum Matter, South China Normal University, Guangzhou}
\affiliation{Guangdong Provincial Key Laboratory of Nuclear Science, Institute of Quantum Matter, South China Normal University, Guangzhou 510006, China}
\affiliation{Guangdong-Hong Kong Joint Laboratory of Quantum Matter, Southern Nuclear Science Computing Center, South China Normal University, Guangzhou 510006, China}

\begin{abstract}
We probe the existence of the $^3_{\Lambda_c}$H where the $N\Lambda_c$ potentials are derived from the quark-delocalization color-screening model (QDCSM).
The $N\Lambda_c$ system is studied and the $N\Lambda_c$ scattering length so as the effective range are obtained in the QDCSM. 
We construct effective Gaussian-type $N\Lambda_c$ potentials
which reproduce the $N\Lambda_c$ scattering data given by the QDCSM.
By solving the $NN\Lambda_c$ three body Schr\"odinger equation
with the Gaussian expansion method, we calculate the energies of the $^3_{\Lambda_c}$H with isospin $I=0$, $J^P=1/2^+$ and $I=0$, $J^P=3/2^+$ under
different color screening parameter $\mu$.
The $J^P=1/2^+$ and $J^P=3/2^+$ states are both bound
when the color screening parameter $\mu$ is set to 1.0 or 1.2, where the $J^P=1/2^+$ 
state is bound by $0.08\sim0.85$ MeV and the $J^P=3/2^+$ state is bound by $0.15\sim1.31$ MeV with respect to the deuteron-$\Lambda_c$ threshold.

\end{abstract}
\maketitle
\section{Introduction}
The low-energy hadronic system and hadron-hadron interactions
are difficult to study due to the non-perturbative nature of Quantum Chromodynamics (QCD). In order to understand the hadron-hadron interactions and exotic hadronic states, various sophisticated models have been applied.
The nucleon-nucleon (NN), hyperon-nucleon (YN) and hyperon-hyperon (YY) interactions have been well constructed by some approaches such as one-boson-exchange (OBE) model~\cite{Machleidt1989,Rijken2006,Rijken2006-2}, chiral effective theory~\cite{Epelbaum2008}, chiral quark model~\cite{Valcarce2005,Fujiwara1995,Zhang1997} and quark-delocalization color-screening model (QDCSM)~\cite{Wang1992,Ping1998,Ping2000}. 
Recently, the hadron-hadron interactions are extended to the heavy flavor systems. For instance, in Refs.~\cite{Wu2010,Wang2011}, the $\Sigma_c\Bar{D}$ bound state are found with chiral quark model. As for the charmonium-nucleon ($c\bar{c}-N$) interactions,
since they have no valence quarks in common, the single meson exchange is suppressed and the single gluon exchange 
is prohibited \cite{hiyama2013}. 
Alternatively, the multi-gluon exchange interactions could be possible as discussed in 
Refs. ~\cite{{Brodsky1997,Kharzeev1994,Luke1992,Brodsky1989}} and the $c\bar{c}-N$ interaction is found to be attractive.

The $\Lambda_cN$ interaction and the possible $\Lambda_cN$ bound state have been studied in several models. With the chiral constituent quark mode,
the S-wave $\Lambda_c$N interaction is investigated in Ref. ~\cite{Gal2014} and they 
find a $J^P=2^+$ candidate of the
$\pi\Lambda_c$N three body system. The attractive force between $\Lambda_c$ and $N$ is also found in Ref.~\cite{Liu2011} within the 
the one meson exchange model and the quark cluster model, where a bound $\Lambda_cN$ state with 1 MeV binding energy is obtained. 
However, the HAL QCD collaboration~\cite{Miyamoto2017} obtained a weakly attractive S-wave $\Lambda_cN$ interaction based
on unphysical quark masses corresponding to pion masses of
$m_\pi= 410–700$ MeV and no $\Lambda_cN$ molecular state is confirmed.  
Recently, an extrapolation of this HAL QCD results to the physical pion mass with the chiral effective filed theory (EFT)~\cite{Epelbaum2003} was presented~\cite{Haidenbauer2017LQCD,Haidenbauer2020LQCD}. 
It gives that the $\Lambda_cN$
interaction at the physical point is slightly
stronger than for that at the large pion masses but still less attractive
than the phenomenological predictions for the $\Lambda_cN$ interaction mentioned above~\cite{Gal2014,Liu2011}.
As a summary, the existence of the $\Lambda_cN$ bound state is unclear and the theoretical study still remains to be continued~\cite{Dover1977}.

Due to the short lifetime of the $\Lambda_c$, it is difficult to directly study the $\Lambda_cN$ interaction from the nuclear scattering. Thus, similar as the hypernuclear physics, people can polish the YN and YY interactions from the spectra data of the light hypernuclei instead of the scattering data. 
For example, in Ref.~\cite{Hiyama2000}, the two different YN spin-orbit interactions derived from the quark model and OBE model give
different $5/2^+-3/2^+$ energy splitting in $^{13}_\Lambda$C or the $3/2^+-1/2^+$ energy splitting in $^{9}_\Lambda$Be. And the observed values~\cite{E929-1,E929-2,E930} 
may suggest which model of YN spin-orbit interaction is favored~\cite{Hiyama2009,Hashimoto2006}.
Therefore, in order to further study the $\Lambda_c$N interaction,
it's a natural consideration to extend our research from the hypernuclei to the charmed hypernuclei. 

In experiment, the first charmed hypernuclei
is claimed to be found in Ref.~\cite{Batusov1981}. 
The experimental search
of the charmed and bottom hypernuclei was also performed
at the ARES facility \cite{Bressani1989}.
Several factories including the $\tau$ charm factory ~\cite{Bunyatov1991} and the kaon factory (such as the Japan Hadron Facility) may have potential to generate the charmed hypernuclei.
In the future, several facilities such as GSI-FAIR,
J-PARC \cite{Riedl2007} or HIAF may be available to check whether the
charmed hypernuclei exist or not. 

In fact, the theoretical study of the $\Lambda_c$ hypernuclei comes back 
to the mid 70s ~\cite{Dover1977,Bando1981,Bando1983,Bando1985,Gibson1983,Bhamathi1981,Tsushima2002a,Tsushima2002b}.
Recently, in Ref.~\cite{Maeda2015}, with the one meson exchange model and the quark cluster model, they found a $NN\Lambda_c$ bound state. Soon after that, Garcilazo et al. 
applied the $\Lambda_cN$ interaction derived from the chiral quark model
to the $J^P=3/2^+$ charmed hypertriton and obtained a bound state with binding energy around
$0.140\sim0.715$ MeV. Recently, Lattice QCD calculation \cite{Miyamoto2017} finds that only the $\rm{A}\ge11$ charmed hypernculei can be bound with a few MeVs binding energy.

As we know, the forces between atoms (molecular force) are qualitative and similarly between the nucleons.
The quark-delocalization color-screening model
(QDCSM), which models the molecular force,
was developed by Wang et al. in 1992 \cite{Wang1992}
and has been extended to various baryons-baryons interactions and scattering
phase shifts ~\cite{Ping1998,Ping2000} in the framework of the resonating group method
(RGM)~\cite{Kamimura1977}.
In QDCSM, quarks confined in one baryon are allowed to
delocalize to another baryon. The delocalization
parameter is determined by the dynamics of the interacting
quark system.
Besides, the color-screening confinement interaction is used
when the two quarks are in different baryons.
This gives an extra parameter $\mu$ which is normally determined  
by the mass of two body hadron-hadron system.

With QDCSM, co-authored by two of the present authors~\cite{Huang2013}, an attractive $\Lambda_cN$ interaction with a repulsive core was obtained and no $\Lambda_cN$ bound state was found. But no charmed hypertriton was investigated. 
Thus in this work, we investigate the possibility of the existence of the charmed hypertriton. With the effective $\Lambda_cN$ interaction
derived from the QDCSM, we calculate the $NN\Lambda_c$ three body system. In order to
solve the three body Schr\"odinger equation, we apply the Gaussian expansion method (GEM)~\cite{Hiyama2003GEM}. 

The paper is organized as follows. In Sec. \uppercase\expandafter{\romannumeral2}, we
introduce the effective $\Lambda_cN$ interaction obtained from the QDCSM.
The detail of GEM is explained in Sec. \uppercase\expandafter{\romannumeral3}
After explaining Method employed,
we show the results and the discussion in Sec.~\ref{sec:results}.  
Summary is at the end.

\section{quark-delocalization color-screening model and $N\Lambda_c$ potential}
In this section, we introduce the QDCSM and its application in the $\Lambda_cN$ system.
In order to describe the $\Lambda_cN$ system, we use the chiral constituent quark model
and the Hamiltonian of the six quarks is:
\begin{equation}
	H=\sum_{i=1}^{6}\left(m_{i}+\frac{p_{i}^{2}}{2 m_{i}}\right)-T_{c}+\sum_{i<j}\left[V^{G}\left(r_{i j}\right)+V^{\chi}\left(r_{i j}\right)+V^{C}\left(r_{i j}\right)\right],
\end{equation}
where $i\;(j)=1\sim6$ is the label of the six quarks composed of the two baryons.
The first term means the sum of the masses and kinetic energies of all the quarks.
$T_c$ is the kinetic energy of the center-of-mass. $V^C$, $V^{G}$ and $V^{\chi}$ indicates the confining potential, gluon-exchange potential and meson-exchange potential where $\chi$ denotes the $\pi$, K or $\eta$ meson, respectively. 
The forms of $V^{G}$ and $V^{\chi}$ are given as follows:
\begin{equation}
\begin{aligned}
V^G\left(r_{i j}\right)&=\frac{ \alpha_{s_{i j}}}{4} \lambda_i \cdot \lambda_j\left[\frac{1}{r_{i j}}-\frac{3}{4 m_i m_j r_{i j}^3} S_{i j} \right.\\
&\left. -\frac{\pi}{2}\left(\frac{1}{m_i^2}+\frac{1}{m_j^2}+\frac{4 \sigma_i \cdot \sigma_j}{3 m_i m_j}\right) \delta\left(r_{i j}\right)\right], \\
V^\chi\left(r_{i j}\right)&=\frac{\alpha_{c h}}{3}\frac{\Lambda^2}{\Lambda^2-m_\chi^2} m_\chi\left\{\left[Y\left(m_\chi r_{i j}\right)-\frac{\Lambda^3}{m_\chi^3} Y\left(\Lambda r_{i j}\right)\right] \sigma_i \cdot \sigma_j \right.\\
&\left.+\left[H\left(m_\chi r_{i j}\right)-\frac{\Lambda^3}{m_\chi^3} H\left(\Lambda r_{i j}\right)\right] S_{i j}\right\} \mathbf{F}_i \cdot \mathbf{F}_j,  \\
\end{aligned}
\end{equation}
where $\lambda_i$ and $F_i$ are the color Gell-Mann matrices and SU(3) falvor matrices, respectively.
The H(x) and Y(x) are the standard Yukawa functions~\cite{Valcarce2005}.
The $S_{ij}$ is the tensor operator:
\begin{equation}
\begin{aligned}
S_{i j}  &=\frac{\left(\sigma_i \cdot \mathbf{r}_{i j}\right)\left(\sigma_j \cdot \mathbf{r}_{i j}\right)}{r_{i j}^2}-\frac{1}{3} \sigma_i \cdot \sigma_j.
\end{aligned}
\end{equation}
For the confining potential, $V^C$, the color screening confining potential is used when
the two quarks are in different baryons:
\begin{equation}
\begin{aligned}
&V^C\left(r_{i j}\right) =-a_c \lambda_i \cdot \lambda_j\left[f\left(r_{i j}\right)+V_0\right], \\
&f\left(r_{i j}\right)  = \begin{cases}r_{i j}^2  &i, j \text { are in the same baryon orbit, } \\
\frac{1-e^{-\mu r_{i j}}}{\mu}  &i, j \text { are in different baryon orbits. }\end{cases} \\
\label{eq:solorscreen}
\end{aligned}
\end{equation}

It should be noted that, in order to determine the color screening parameter $\mu$, we need the total mass of the two baryon system as an input. 
In the $NN$ system, the parameter $\mu=1.0$ is used to fit deuteron mass. 
In this work, we treat $\mu$ as a relative common parameter since we do not have the $N\Lambda_c$ mass. 
The other parameters are determined with a global fitting to the masses of
light flavor mesons and charmed baryons $\Lambda_c$, $\Sigma_c$ and $\Sigma^*_c$~\cite{Chen2011,Huang2013}.

Normally, the usual quark cluster model gives the single quark wave function as follows:
\begin{equation}
\begin{aligned}
\phi_\alpha\left(\mathbf{S}_i\right) & =\left(\frac{1}{\pi b^2}\right)^{3 / 4} e^{-\frac{1}{2 b^2}\left(\mathbf{r}_\alpha-\mathbf{S}_i / 2\right)^2}, \\
\phi_\beta\left(-\mathbf{S}_i\right) & =\left(\frac{1}{\pi b^2}\right)^{3 / 4} e^{-\frac{1}{2 b^2}\left(\mathbf{r}_\beta+\mathbf{S}_i / 2\right)^2} 
\end{aligned}
\end{equation}
where $r_{\alpha(\beta)}$ is the coordinate of the single quark to the center of the baryon and $S_i$ is the generating coordinate between the two baryons.
In QDCSM, the delocalized quark wave functions are used:
\begin{equation}
\begin{aligned}
\psi_\alpha\left(\mathbf{S}_i, \epsilon\right) & =\left(\phi_\alpha\left(\mathbf{S}_i\right)+\epsilon \phi_\alpha\left(-\mathbf{S}_i\right)\right) / N(\epsilon), \\
\psi_\beta\left(-\mathbf{S}_i, \epsilon\right) & =\left(\phi_\beta\left(-\mathbf{S}_i\right)+\epsilon \phi_\beta\left(\mathbf{S}_i\right)\right) / N(\epsilon), \\
N(\epsilon) & =\sqrt{1+\epsilon^2+2 \epsilon e^{-S_i^2 / 4 b^2}}, \\
\end{aligned}
\end{equation}
where $\epsilon(S_i)$ is determined
by the dynamics of the quark system.

\begin{table}[tbh]
\caption{The channels coupled to the $N \Lambda_c$ states.}
\begin{tabular}{p{2.0cm}<{\centering}p{6.0cm}<{\centering}}
  \hline\hline
     States     & Channels     \\
  \hline
 $J^{P}=0^{+}$    & $N \Lambda_c\left({ }^1 S_0\right), N \Sigma_c\left({ }^1 
                  S_0\right), N \Sigma_c^*\left({ }^5 D_0\right)$      \\
 \hline
 $J^{P}=1^{+}$ & $N \Lambda_c\left({ }^3 S_1\right), N \Sigma_c\left({ }^3 S_1\right), N \Sigma_c^*\left({ }^3 S_1\right)$, \\
& $N \Lambda_c\left({ }^3 D_1\right), N \Sigma_c\left({ }^3 D_1\right), N \Sigma_c^*\left({ }^3 D_1\right), N \Sigma_c^*\left({ }^5 D_1\right)$ \\
\hline
\end{tabular}
\label{tab:NLc-channels}
\end{table}

In Ref.~\cite{Huang2013}, the $J^{P}=0^+$ and $J^{P}=1^+$ $N\Lambda_c$ scattering phase shifts and the potentials have been studied in QDCSM with the well-developed RGM. 
The coupled channel effects are considered in the $N\Lambda_c$ system. 
In total, three/seven channels are included in the  $J^{P}=0^+$/$J^{P}=1^+$ $N\Lambda_c$ channel (shown in Tab.~\ref{tab:NLc-channels}). 

\begin{figure}[htbp]
\setlength{\abovecaptionskip}{0.cm}
\setlength{\belowcaptionskip}{-0.cm}
\centering
\includegraphics[width=0.46\textwidth]{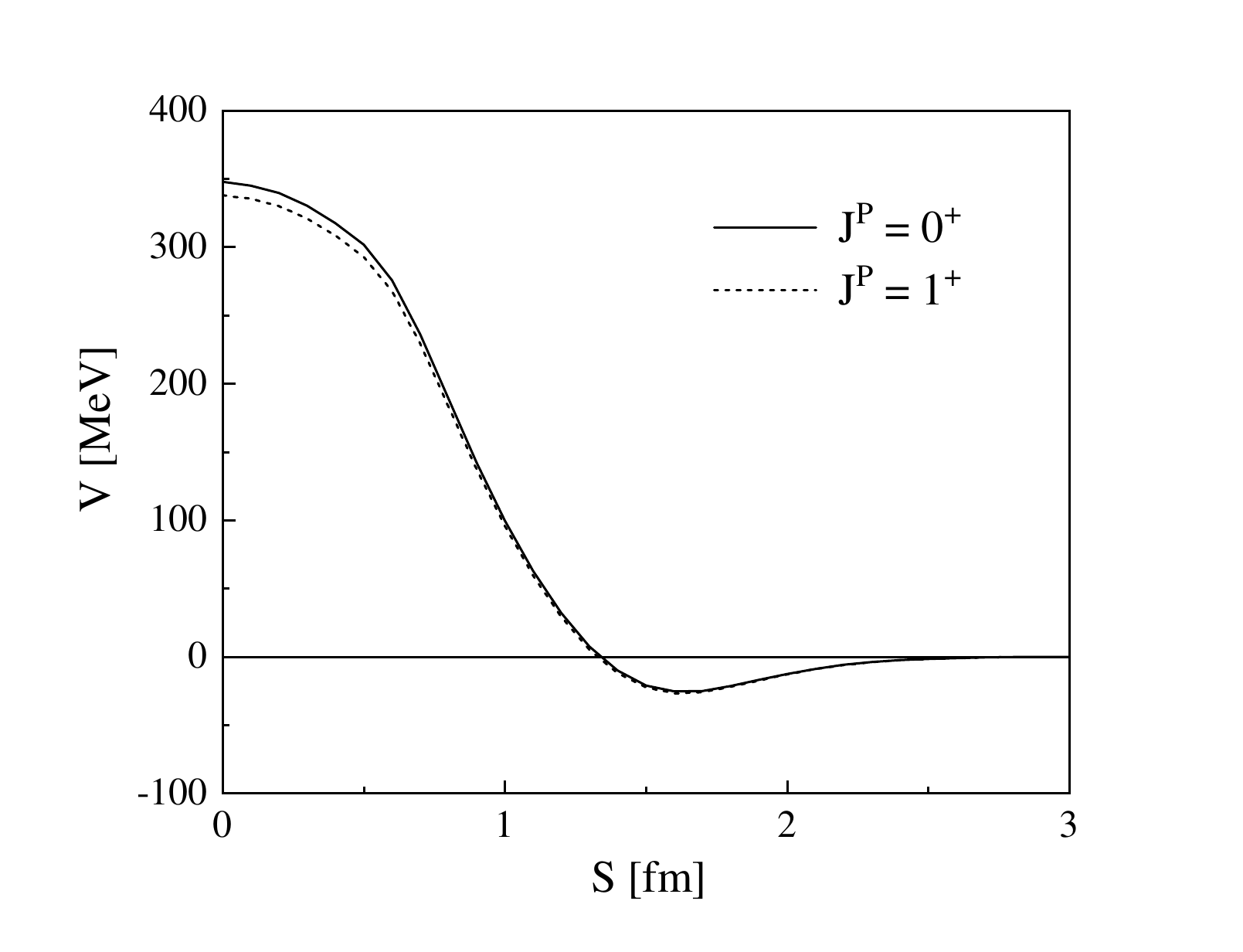}
\caption{Normalized $N\Lambda_c$ potentials with respect to the generating coordinates S}
\label{fig:V(S).PDF}
\end{figure}
In this work, we diagonalize the nondiagonal S matrix and normalize the coupled channel effects into the single channel.
We normalize the effects of all three channels into $N \Lambda_c\left({ }^1 S_0\right)$
in the $J^{P}=0^+$ case. Similarly in the $J^{P}=1^+$ state, the effects of all seven channels are normalized into $N \Lambda_c\left(^3 S_1\right)$ channel. Thus, the $\Sigma_c\;(\Sigma^*_c)$ and the D-wave components are normalized to the S-wave $N\Lambda_c$ potential.
In Fig.~\ref{fig:V(S).PDF}, we depict the normalized the $J^{P}=0^+$ and $J^{P}=1^+$ $N\Lambda_c$ potentials where $\mu=1.0$.

We want to stress that the $N\Lambda_c$ potentials in Fig.~\ref{fig:V(S).PDF} 
are constructed with respect to the generating coordinate $S_i$, which is not equivalent to the relative distance $r$ between the center-of-mass of the two baryons. The generating coordinate $S_i$
is constructed based on the reference center of the three quarks, which is not the center-of-mass of the baryons (N or $\Lambda_c$) especially when the two baryons are close and the
quark delocalization effect is strong.
Then, the difference between the V(S) and V(r) in the short-range could be large.
Therefore, it is not a proper way to directly apply this potential into the $NN\Lambda_c$ three body system. On the other hand, if we study
the $NN\Lambda_c$ system within framework of the QDCSM, a nine quarks system is necessary to solve, which is not practical with our present capacity. Instead, we construct the effective S-wave $\Lambda_cN$ interactions which reproduce the $\Lambda_cN$ scattering length ($a$) and the effective range $r_0$ defined as follows: 
\begin{equation}
\begin{aligned}
k\cot{\delta}=\frac{1}{a}+\frac{1}{2}r_0k^2+..., 
\label{eq:delta}
\end{aligned}
\end{equation}
where $\delta$ is the scattering phase shift and $k$ is the wave number.
The values are shown in Tab.~\ref{tab:scatteringlength} for
three different values, i.e. 0.8, 1.0 and 1.2, of $\mu$.

\begin{table}[tbh]
\caption{The scattering length and the effective range of $J^{P}=0^+$ and $J^{P}=1^+$ $N \Lambda_c$ systems.}
\begin{tabular}{p{2.0cm}<{\centering}p{3.0cm}<{\centering}p{3.0cm}<{\centering}}
  \hline\hline
     $(J^{P},\mu)$   &  Scattering length (fm) & Effective range (fm)    \\
  \hline
 $0^{+},0.8$   & -1.57 & 5.06 \\
 $0^{+},1.0$   & -4.88 & 3.62 \\
 $0^{+},1.2$   & -2.58 & 3.65 \\ 
 \hline
 $1^{+},0.8$   & -1.92 & 3.70 \\
 $1^{+},1.0$   & -7.11 & 3.15 \\
 $1^{+},1.2$   & -2.88 & 3.50 \\ 
\hline
\end{tabular}
\label{tab:scatteringlength}
\end{table}

\begin{figure}[htbp]
\setlength{\abovecaptionskip}{0.cm}
\setlength{\belowcaptionskip}{-0.cm}
\centering
\includegraphics[width=0.46\textwidth]{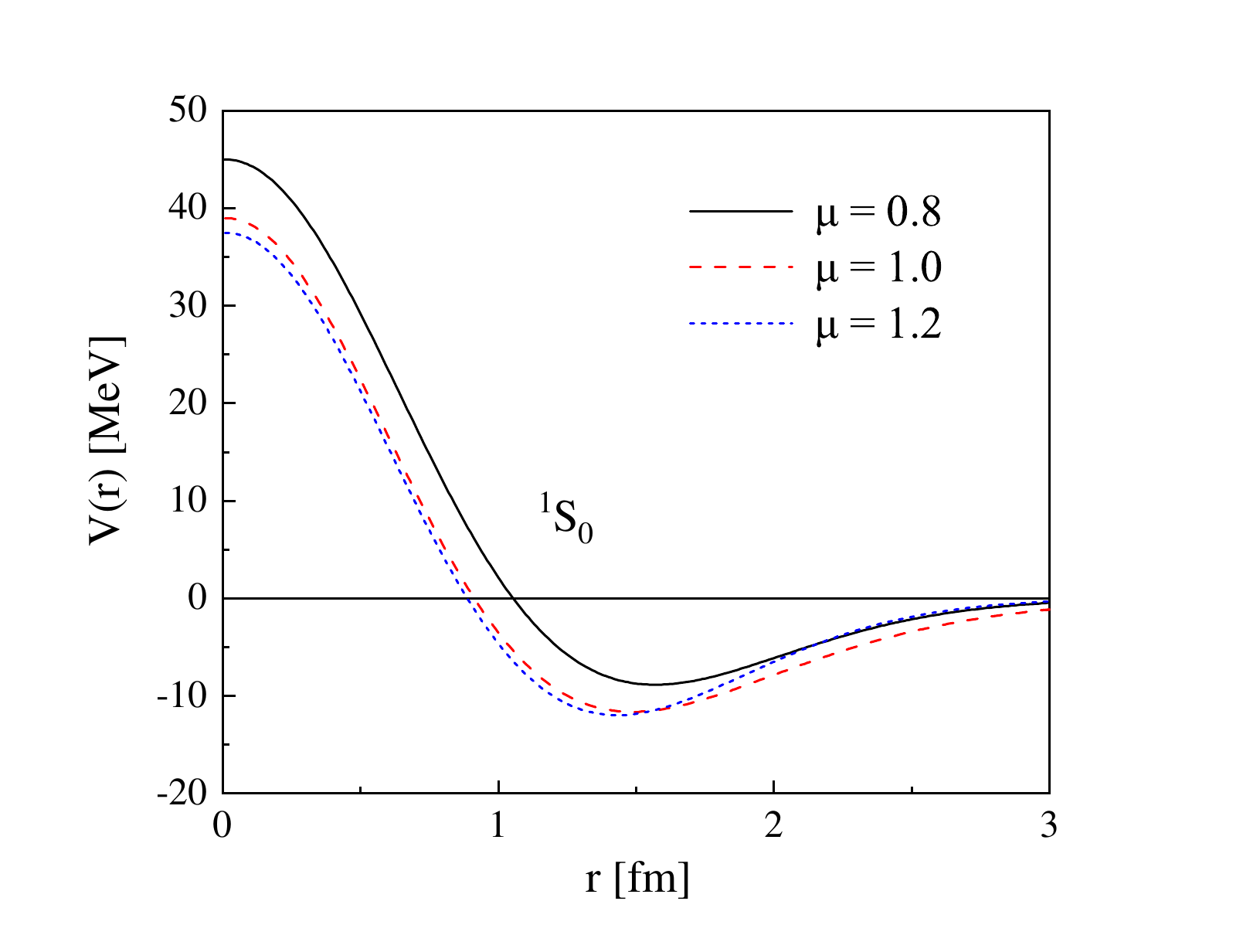}
\caption{Effective $N\Lambda_c$ interaction of $^1S_0$ case.}
\label{fig:Vqdcsm0}
\end{figure}

\begin{figure}[htbp]
\setlength{\abovecaptionskip}{0.cm}
\setlength{\belowcaptionskip}{-0.cm}
\centering
\includegraphics[width=0.46\textwidth]{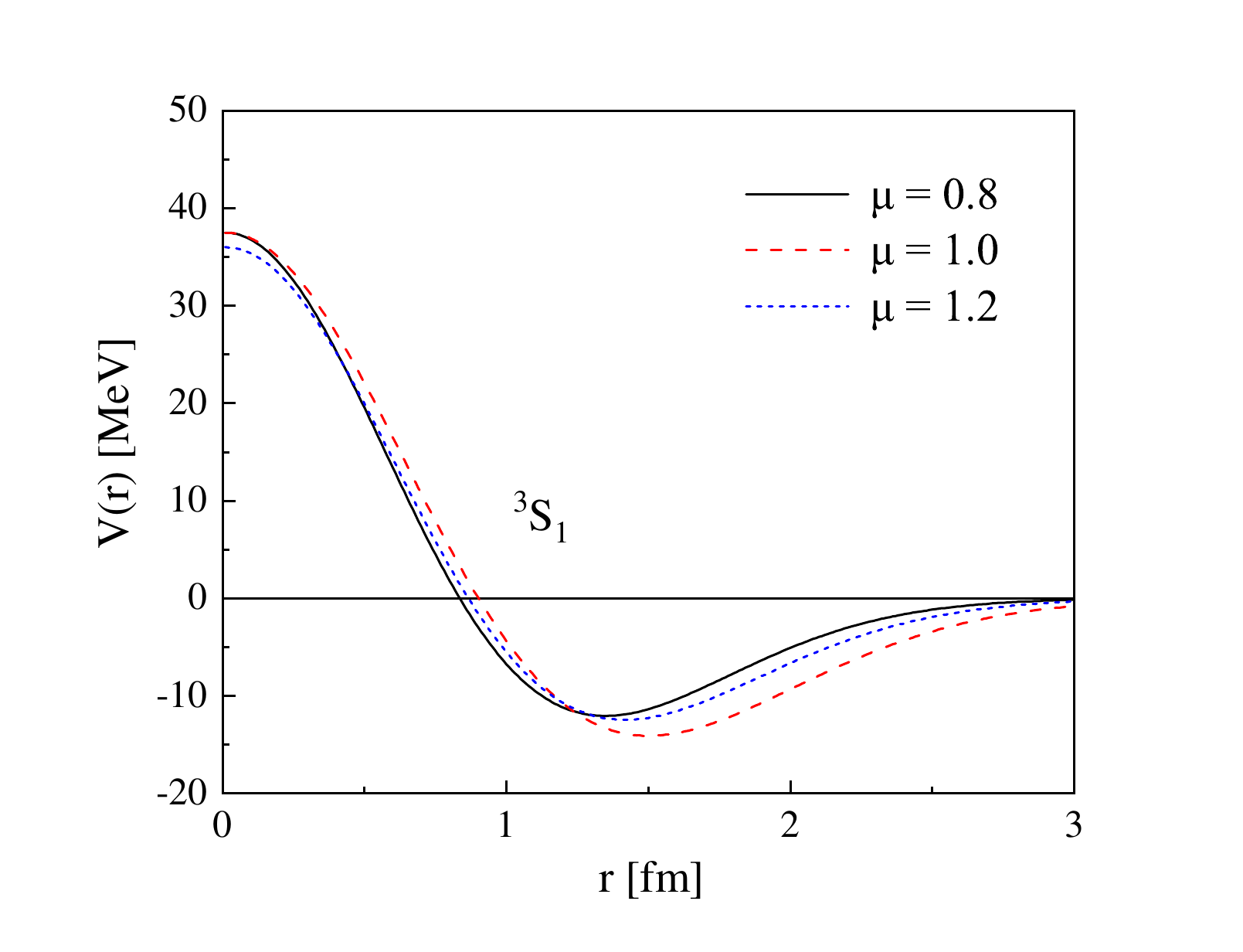}
\caption{Effective $N\Lambda_c$ interaction of the $^3S_1$ case.}
\label{fig:Vqdcsm1}
\end{figure}

The functional form of the effective $N\Lambda_c$ interactions are given by:
\begin{equation}
V_{\mathrm{eff}}(r)=V_1e^{-r^2/b_1^2}+V_2e^{-r^2/b_2^2}.
\label{eq:veff}
\end{equation}
The fitting parameters are listed in Tab.~\ref{tab:para}.
The obtained effective $\Lambda_cN$ interactions are shown in Fig.~\ref{fig:Vqdcsm0} and Fig.~\ref{fig:Vqdcsm1}, respectively for the $J^{P}=0^+$ and $J^{P}=1^+$ channel.
As we can see, the effective potential V(r) and the potential V(S) shown in Fig.~\ref{fig:V(S).PDF} have a significant difference in the short range of the r (or S), which consists with our conjectures that the quark delocalization effect is strong when the two baryons are close.

\begin{table}[tbh]
\caption{Fitting parameters of $V_{\mathrm{eff}}(r)$ defined in Eq.~\ref{eq:veff} for the $N\Lambda_c$ effective potentials. The $V_1$ and $V_2$ are in unit of MeV. $b_1$ and $b_2$ are in unit of fm.}
\begin{tabular}{p{0.8cm}<{\centering}|p{1.1cm}<{\centering}p{1.1cm}<{\centering}p{1.1cm}<{\centering}|p{1.1cm}<{\centering}p{1.1cm}<{\centering}p{1.1cm}<{\centering}}
  \hline\hline
    & \multicolumn{3}{c|}{$^1S_0$}  & \multicolumn{3}{c}{$^3S_1$}   \\
  \hline
   $\mu$ & 0.8 & 1.0 &1.2  & 0.8 & 1.0 &1.2  \\
   \hline
   $V_1$  & -115.5 & -42.0 & -223.5  & 255.0 & -216.4 & -226.5  \\
    $b_1$ & 1.29 & 1.58 &1.20  & 1.12 & 1.30 &1.20  \\
    $ V_2 $ & 160.6 & 81.3 & 261.2  & 292.8 & 253.3 & 262.6  \\
      $b_2$  & 1.05 & 0.91 &1.05  & 1.01 & 1.12 &1.05  \\
\hline
\end{tabular}
\label{tab:para}
\end{table}


\section{Gaussian Expansion method}
\begin{figure}[htbp]
\setlength{\abovecaptionskip}{0.cm}
\setlength{\belowcaptionskip}{-0.cm}
\centering
\includegraphics[width=0.46\textwidth]{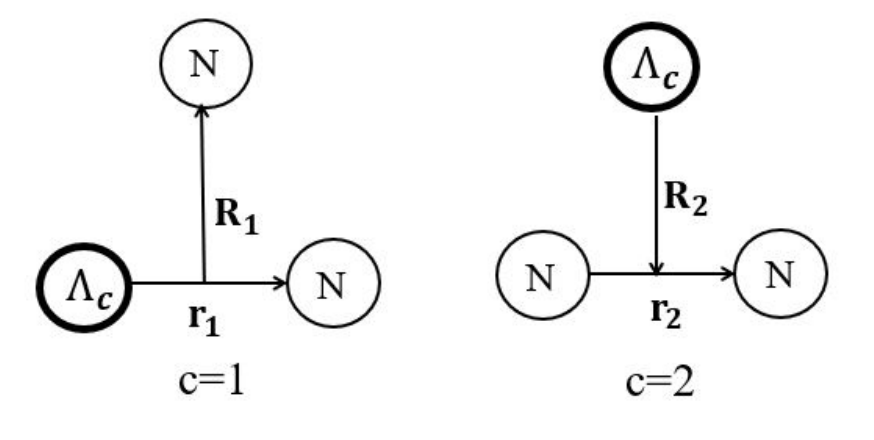}
\caption{Jacobi coordinates of NN$\Lambda_c$ three body system.}
\label{fig:NNLcjaco.PDF}
\end{figure}

In order to solve the $NN\Lambda_c$ three-body Schr\"odinger
equation, we apply the Gaussian expansion method and set the three body wave function as:
\begin{equation}
\begin{aligned}
\Psi_{JMTT_z}\left(_{\Lambda_c}^3\rm{H}\right)&= \sum_{c=1}^{2} \sum_{s,S,L}\sum_{n_1,l_1}\sum_{n_2,l_2}C_{
 \gamma}^{(c)}\mathcal{A} \\
&\times\left\{\left\{\left(\phi_{n_1 l_1}^{(c)}\left(\boldsymbol{r}_{c}\right) \psi_{n_2 l_2}^{(\mathrm{c})}\left(\boldsymbol{R}_{c}\right)\right)_{L} \right.\right. \\
&\left.\times\left[\left(\chi^{1}_{1/2}\chi^{2}_{1/2}\right)_s\chi^{3}_{1/2}\right]_S\right\}_{JM} 
\left.\left(\tau^{1}_{1/2}\tau^{2}_{1/2}\right)_{TT_z}\right\}
, \label{eq:wf1}
\end{aligned}
\end{equation}
where the two sets of Jacobian coordinates (labeled as 1 and 2) are shown in Fig.~\ref{fig:NNLcjaco.PDF}. Here $\gamma$ denotes $\{L,s,S,n_1,l_1,n_2,l_2\}$.
$\chi$ and $\tau$ represent the spin and isospin wave function of the nucleon or
$\Lambda_c$, respectively. 
Note that we omit the isospin wave function of the $\Lambda_c$ since it is isospin singlet. 
The $\mathcal{A}$ is the anti-symmetric operator between the two nucleons.
The relative wave functions between the baryons, corresponding to the two Jacobi coordinates, $\phi_{n_1 l_1}(\boldsymbol{r})$, $\psi_{n_2 l_2}(\boldsymbol{R})$ are expanded by using the following Gaussian basis functions, applying the GEM:
\begin{eqnarray}
&\phi_{n_1 l_1}(\boldsymbol{r})=r^{\ell_1} e^{-(r/r_{n_1})^2}Y_{\ell_1 m_1}(\hat{\vc r}), \nonumber \\
&\psi_{n_2 l_2}(\boldsymbol{R})=R^{\ell_2} e^{-(R/R_{n_2})^2}Y_{\ell_2 m_2}(\hat{\vc R}). 
\end{eqnarray}
The Gaussian variational parameters are chosen to have geometric progression:
\begin{eqnarray}
&r_{n_1}=r_{\rm min} A_1^{n_1-1},  \quad (n_1=1 \sim n_1^{\rm max}), \nonumber \\
&R_{n_2}=R_{\rm min} A_2^{n_2-1},  \quad (n_2=1 \sim n_2^{\rm max}). 
\label{eq:gem2}
\end{eqnarray}
Then, the eigen energies and the coefficients $C_{\gamma}$ are obtained with applying the Rayleigh-Ritz variational method.

\section{Results and Discussions}
\label{sec:results}
\begin{figure*}[tb]
\setlength{\abovecaptionskip}{0.cm}
\setlength{\belowcaptionskip}{-0.cm}
\centerline{
\includegraphics[width=14.0 cm,height=10.0 cm]{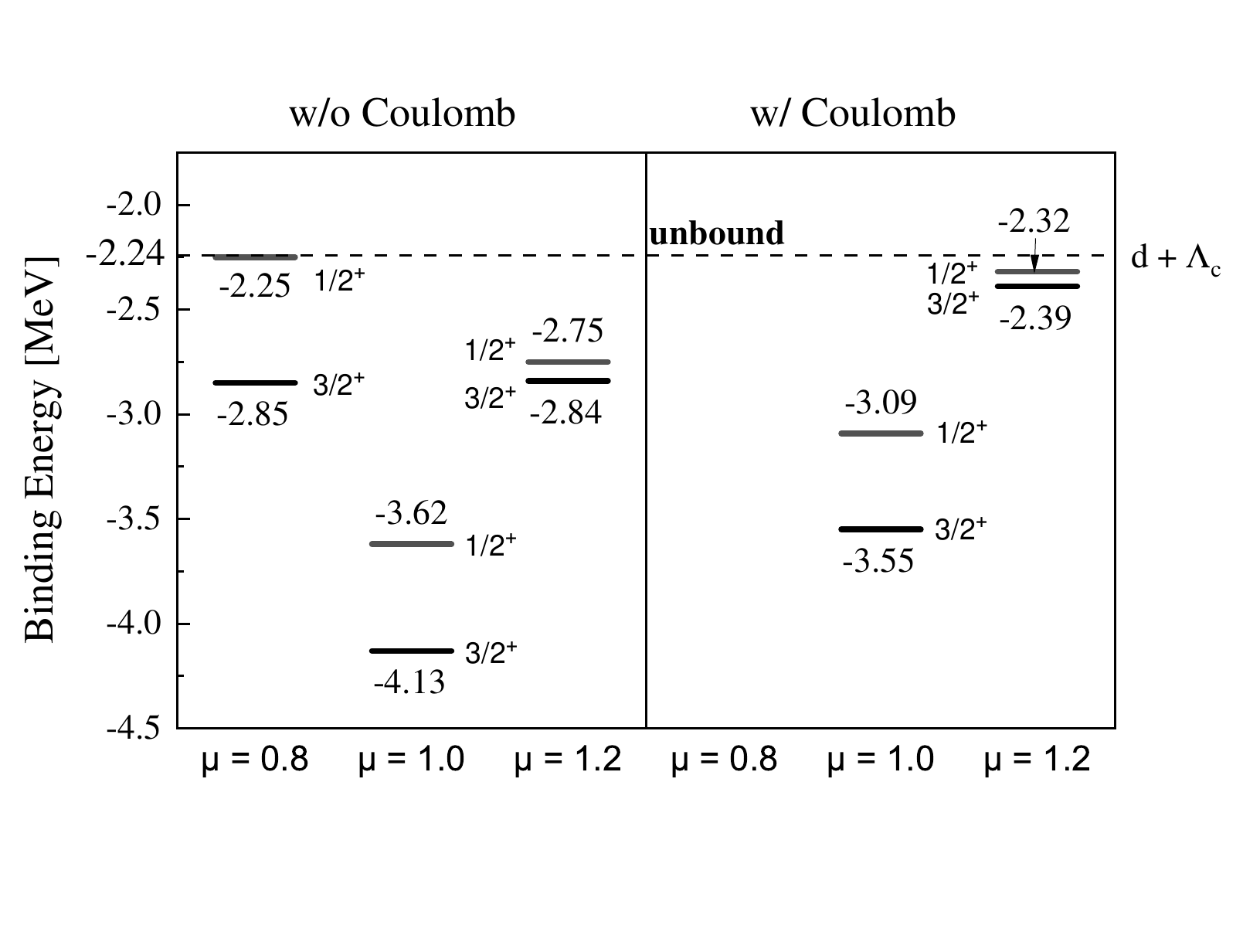}}
\caption{The calculated energy levels of $^3_{\Lambda_c}$H. The levels
without and with considering the coulomb force are shown in the left figure and the right figure, respectively, for the color screening parameter $\mu=0.8,1.0,1.2$. The energies are given with respect to the $NN\Lambda_c$ three body breakup threshold. The $d+\Lambda_c$ threshold is also shown at $E=-2.24$ MeV. 
}
\label{fig:0pspec}
\end{figure*}

The energy levels of $^3_{\Lambda_c}$H with $I=0$, $J^P=1/2^+$ and $J^P=3/2^+$ states
are shown in Fig.~\ref{fig:0pspec}. The binding energies of $^3_{\Lambda_c}$H 
are given with respect to
the $N+N+\Lambda_c$ three body break-up threshold.
We calculate the energies of $^3_{\Lambda_c}$H without or with the coulomb force between the $\Lambda_c^+$ and the nucleon, which are shown in the left and right column
of Fig.~\ref{fig:0pspec}.
As shown in Fig.~\ref{fig:0pspec}, it is obvious that the binding energies are smaller when we include the repulsive coulomb force.

The binding energy for the $\mu=1.0$ case is larger than the $\mu=1.2$ and $\mu=0.8$ cases.
This can be expected from the $N\Lambda_c$ scattering length in Table.~\ref{tab:scatteringlength}, in which the $\mu=1.0$ case has the largest scattering length for both $^1S_0$ and $^3S_1$.  
In the $\mu=0.8$ case, no bound state is obtained with the coulomb force.
And when we exclude coulomb force, a very weakly bound state for the
$1/2^+$ state is obtained with the $\Lambda_c$ separation energy $B_{\Lambda_c}=0.01$ MeV.
As for the $\mu=1.2$ case, very weakly bound states with $B_{\Lambda_c}=0.08$ MeV and
0.15 MeV are obtained, respectively for the $1/2^+$ and $3/2^+$ states. 

On the other hand, the energy splitting between the $1/2^+$ and $3/2^+$ states does not change significantly when we include the coulomb force since the coulomb force does not have a spin-dependent term.
The $\mu=1.2$ case has the smallest energy splitting while the $\mu=0.8$
and 1.0 cases are similar.

We find that the energy levels of $J^{P}=3/2^+$ state are lower than the $J^{P}=1/2^+$ state in all of the three cases of $\mu$. 
In order to study this, we calculate the fraction of the S-wave wave function (L=0 and S=3/2) in the $J^{P}=3/2^+$ state which is around $94\%$
and the S-wave component (L=0 and S=1/2) is around $97\%$
for the $J^{P}=1/2^+$ state. 
Thus, the $^1S_0$ $N\Lambda_c$ component is nearly excluded in the
$J^{P}=3/2^+$ state while it has a dominant contribution in the
$J^{P}=1/2^+$ state. Besides, the $^3S_1$ $N\Lambda_c$ interaction is more attractive than the $^1S_0$ case. Therefore, the binding energy of the $J^{P}=3/2^+$ state are larger than the $J^{P}=1/2^+$ state.

It's interesting that the similar conclusions were made
in Refs.~\cite{Garcilazo2015,Maeda2015} for the
three body $^3_{\Lambda_c}$H calculations. 
As shown in Fig. 9 of Ref.~\cite{Maeda2015}, the
$J^{P}=1^+$ interaction is also stronger than the $J^{P}=0^+$ case though their calculated $N\Lambda_c$ interaction is much stronger than ours.
For the three body $^3_{\Lambda_c}$H calculation, the binding energy of the $J^{P}=3/2^+$ state is larger than the $J^{P}=1/2^+$ state. In Ref.~\cite{Garcilazo2015}, they also found that the binding energy for the $J^{P}=3/2^+$ state in $^3_{\lambda_c}$H
is larger than the $J^{P}=1/2^+$ state.
\begin{table}[tbh]
\caption{The $\Lambda_c$ separation ($B_{\Lambda_c}$) energy of $^3_{\Lambda_c}$H under different models.
The units are in MeV.}
\begin{tabular}{p{2.0cm}<{\centering}p{1.3cm}<{\centering}p{1.3cm}<{\centering}
p{1.3cm}<{\centering}p{1.7cm}<{\centering}}
  \hline\hline
        &  $\mu=1.0$ &  $\mu=1.2$ & Ref.~\cite{Maeda2015} & Ref.~\cite{Garcilazo2015}  \\
  \hline
 $B_{\Lambda_c}(1/2^+)$  & 0.85  & 0.08 & 18.23 & - \\
 $B_{\Lambda_c}(3/2^+)$  & 1.31  & 0.15 & 18.89 & $0.14\sim0.75$\\
  $1/2^{+}-3/2^{+}$            & 0.46  & 0.07 & 0.66  & -\\ 
\hline
\end{tabular}
\label{tab:energycompare} 
\end{table}
The comparison with those two works~\cite{Garcilazo2015,Maeda2015} are collected in Tab.~\ref{tab:energycompare}, where we show 
the $\Lambda_c$ separation energies
for the $J^{P}=1/2^+$ and $J^{P}=3/2^+$ states
in the $\mu=1.0$ and $\mu=1.2$ cases, as well as the energy difference between the $1/2^+$ and
$3/2^+$. In our calculation, the $\Lambda_c$ separation energies in $\mu=1.2$ case
and the energy of $1/2^+-3/2^+$ are smaller than the $\mu=1.0$ case. 
Despite the extremely large binding energies in Ref.~\cite{Maeda2015}, the $1/2^+-3/2^+$
energy is close to our $\mu=1.0$ result. As for the $\Lambda_c$ separation energies of the $J^P=3/2^+$ state, our results
are consistent with the values in Ref.~\cite{Garcilazo2015}. 

As people know~\cite{exp-hypernuclei}, the lightest hypernucleus, $^3_\Lambda$H\;($1/2^+$), is a weakly bound state and
the $B_\Lambda=0.13\pm0.05$ MeV. And according to our model, a charmed hypertrion may also be bound with the $B_{\Lambda_c}=0.08\sim0.85$ MeV for the $1/2^+$ state.
Though both the hypertriton and the $^3_\Lambda$H($1/2^+$) might have small binding energies,
the dynamical behaviors between the $\Lambda$ and $\Lambda_c$ could be different significantly
in the nucleus due to the much heavier mass of the $\Lambda_c$ baryon.
In order to investigate this issue, we study the three body $^3_\Lambda$H wave functions with calculating the following form: 
\begin{equation}
\begin{aligned}
\rho(R_2)=\int  |\Psi(^3_{\Lambda_c}{\rm H})|^2d^3{\bm{r}}_2d^2\hat{\bm{R}}_2 , \\
\rho(r_2)=\int  |\Psi(^3_{\Lambda_c}{\rm H})|^2d^3{\bm{R}}_2d^2\hat{\bm{r}}_2 ,
\label{eq:den}
\end{aligned}
\end{equation}
where $\emph{R}_2$ and $\emph{r}_2$ are the Jacobian coordinates with channel c=2, defined in Fig.~\ref{fig:NNLcjaco.PDF} and $\Psi(^3_{\Lambda_c}{\rm H})$ is the wave function of the $^3_{\Lambda_c}$H. 

\begin{figure}[htbp]
\setlength{\abovecaptionskip}{0.cm}
\setlength{\belowcaptionskip}{-0.cm}
\centering
\includegraphics[width=0.46\textwidth]{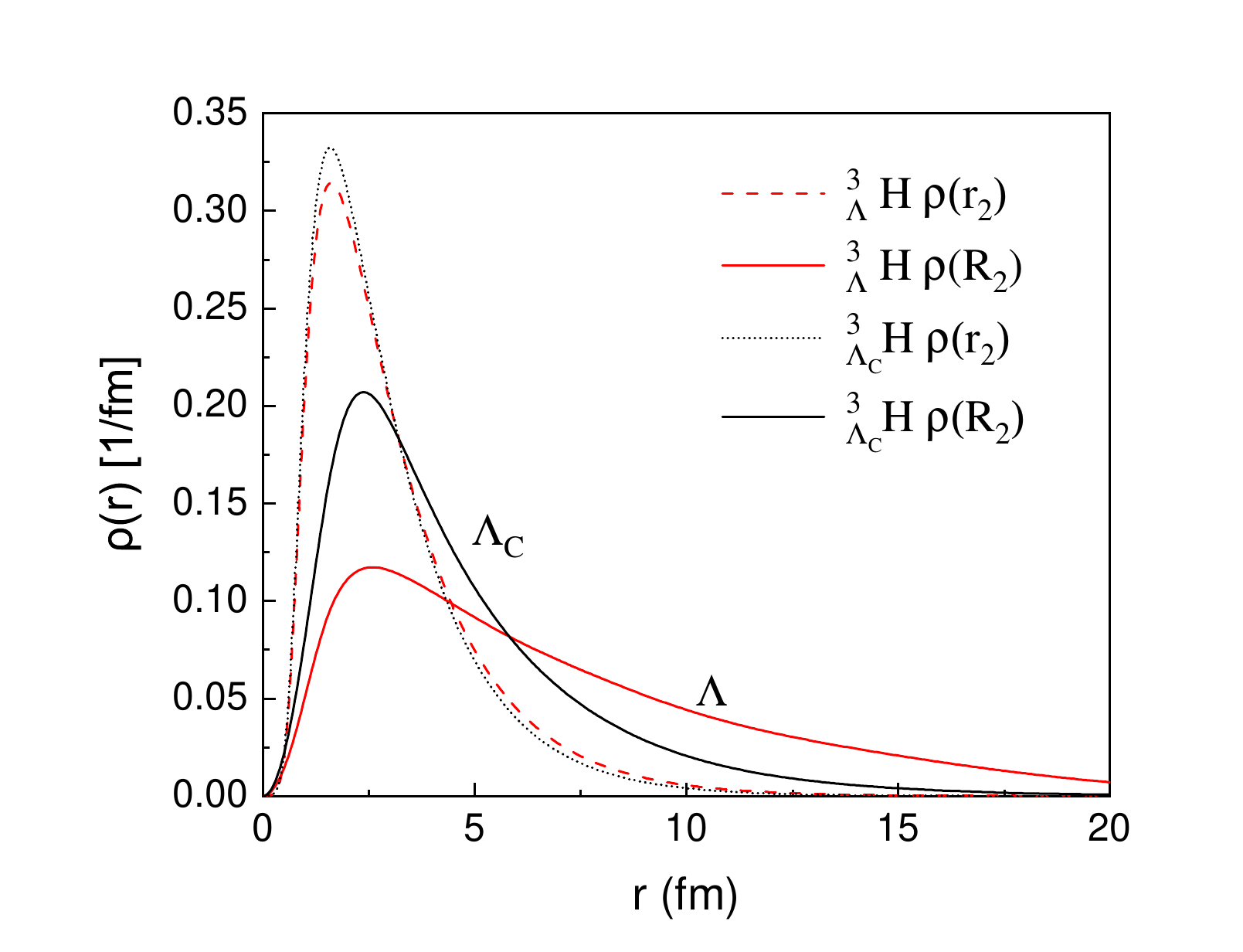}
\caption{$\rho(r_2)$ and $\rho(R_2)$ defined in Eq.~\ref{eq:den}. The
black ones are the $^3_{\Lambda_c}$H system and the red ones are
the hypertriton.
}
\label{fig:density}
\end{figure}

In Fig.~\ref{fig:density}, we depict the $\rho(r_2)$ and $\rho(R_2)$ of the $^3_{\Lambda_c}$H(\;$1/2^+$) state in the $\mu=1.0$ case. It's obvious that
$\rho(r_2)$ reflects the dynamical behavior between the two nucleons and
$\rho(R_2)$ reflects the $\Lambda_c$'s behavior. For comparison, we also
show the $\rho(r_2)$ and $\rho(R_2)$ of the hypertriton. The calculated $B_\Lambda$ is 0.13 MeV which agrees with the experimental data. The NN interactions used in the
hypertriton and $^3_{\Lambda_c}$H($1/2^+$) are the same and we also apply the
GEM to solve the three body $NN\Lambda$ problem. The $\Lambda$N interaction used here is same as the one used in Ref.~\cite{wu2019}
which simulates the G-matric (YN) interaction derived from the Nijmegen model F (NF).
Then, as shwon in Fig.~\ref{fig:density}, we find that the wave function of $\Lambda_c$ is more compact in the $^3_{\Lambda_c}$H(\;$1/2^+$) state than
the $\Lambda$ in the hypertriton though they have similar binding energies. On the other hand, the nucleon-nucleon wave functions are similar with each other in the $^3_{\Lambda_c}$H($1/2^+$) and the hypertriton.
It's well known that in the hypertrion, the $\Lambda$ wave function is very dilute and is around 10 fm away from the deuteron core.
But such behavior may not happen for the $\Lambda_c$ in the charmed hypertrion according to our calculation. 

\section{Summary}
In this work, we investigate the three body $NN\Lambda_c$ system within 
the framework of the quark-delocalization color-screening model and probe the possibility of a bound $^3_{\Lambda_c}$H ($T=0$).

First, with QDCSM, we normalize the $\Sigma\;(\Sigma^*)$ and the $l=2$ components into the S-wave $N\Lambda_c$ constituents. Considering the differences between the generating coordinates $S$ and the relative coordinate $r$ between the baryons, we construct effective Gaussian type $\Lambda_cN$
potentials with reproducing the $\Lambda_cN$ scattering lengths and the effective ranges for the $^1S_0$ and $^3S_1$ cases. As for
the $NN$ interaction, we apply the AV8' realistic $NN$ interaction.

With the above potentials, we solve the three body problem with GEM and obtain the energy levels of $^3_{\Lambda_c}$H $J^P=1/2^+$ and $J^P=3/2^+$ states with different color screening parameters $\mu$. We find that both  
$J^P=1/2^+$ and $J^P=3/2^+$ states are bound under the $\mu=1.0$ and
$\mu=1.2$ cases while it's unbound when $\mu=0.8$. Our calculated
binding energy of $J^P=3/2^+$ state consists with that of Ref.~\cite{Garcilazo2015}. We also find that the
energy difference of $1/2^{+}-3/2^{+}$ is $70\sim 460$ keV and close to another calculation
in Ref.~\cite{Maeda2015}, i.e. 660 keV. 
At the end, we also compare the dynamical behavior of the $\Lambda_c$ in the
$^3_{\Lambda_c}$H and the $\Lambda$ in the hypertrion. We find that the 
$\Lambda_c$ is closer to the nucleon core than the $\Lambda$ in hypertrion due to the $\Lambda_c$'s
heavier mass.frr
The experimental search of the charmed hypertriton is urgently needed in order to further study the charmed hypernuclei.

\begin{acknowledgments}
This work is supported by the Strategic Priority Research Program of Chinese Academy of Sciences under the Grant NO. XDB34030301, the National Natural Science Foundation of China under the Grant NOs. 12005266, 12075288, 11735003, 11961141012, 12035007, Guangdong Provincial funding with Grant No.~2019QN01X172, and Guangdong Major Project of
Basic and Applied Basic Research No. 2020B0301030008. It is also
supported by the Youth Innovation Promotion Association CAS.
Q.W. is also supported by the NSFC and the Deutsche Forschungsgemeinschaft (DFG, German
Research Foundation) through the funds provided to the Sino-German Collaborative
Research Center TRR110 ``Symmetries and the Emergence of Structure in QCD"
(NSFC Grant No. 12070131001, DFG Project-ID 196253076-TRR 110).

\end{acknowledgments}

\end{document}